\documentclass[prb,amsmath,twocolumn,showpacs]{revtex4}
\usepackage{graphicx}
\newcommand{\vect}[1]{\boldsymbol{#1}}
\newcommand{\ddif}[2]{\frac{\mathrm{d}#1}{\mathrm{d}#2}}
\newcommand{\dpar}[2]{\frac{\partial #1}{\partial #2}}
\newcommand{\eff}{\text{eff}}
\newcommand{\abs}[1]{\lvert #1 \rvert}

\begin{document}

\title{Minimal field requirement in precessional magnetization
  switching}

\author{Di Xiao}
\email{dxiao@physics.utexas.edu}
\affiliation{Department of Physics, The University of Texas at Austin,
  Austin, TX 78712}
\author{M. Tsoi}
\affiliation{Department of Physics, The University of Texas at Austin,
  Austin, TX 78712}
\author{Qian Niu}
\affiliation{Department of Physics, The University of Texas at Austin,
  Austin, TX 78712}

\begin{abstract}
We investigate the minimal field strength in precessional
magnetization switching using the Landau-Lifshitz-Gilbert equation in
under-critically damped systems.  It is shown that precessional
switching occurs when localized trajectories in phase space become
unlocalized upon application of field pulses.  By studying the
evolution of the phase space, we obtain the analytical expression of
the critical switching field in the limit of small damping for a
magnetic object with biaxial anisotropy.  We also calculate the
switching times for the zero damping situation.  We show that
applying field along the medium axis is good for both small field and
fast switching times.
\end{abstract}


\maketitle

\section{\label{sec:intro}Introduction} 

Magnetization reversal in magnetic particles and thin films has been a
continuous and growing topic in the past several decades, motivated by
its great application potential in magnetic data storage and random
access memories (RAM).  Recent development of new fabrication
techniques has made it possible to produce nanometer-sized magnetic
objects with well-controlled shape, structure, and chemical
composition.  Magnetic anisotropy of these objects, including shape
anisotropy and magneto-crystalline anisotropy, makes the dynamic
magnetization processes highly nonlinear.  A thorough understanding of
micromagnetic dynamics is thus desirable as many efforts have already
been made in this field.\cite{spindynamicsi, spindynamicsii}

\begin{figure}[b]
\includegraphics[width=8cm]{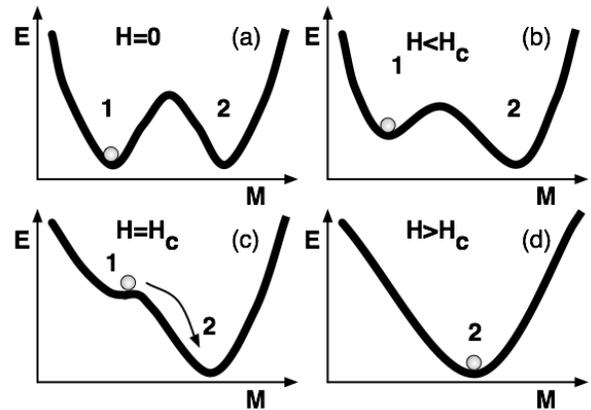}
\caption{\label{fig:landscape}Typical energy landscapes: energy $E$ vs
  magnetization $M$ at different value of applied field $H$. In (a)
  $H=0$, system sits on one of its local minimum 1 separated with 2 by
  a barrier.  In (b) $H < H_c$, the critical field, position 1 is
  still a local minimum and magnetization reversal will not occur.  In
  (c) $H=H_c$, the minimum 1 and the barrier coalesce, any fluctuation
  will cause the system jump to minimum 2.  In (d) $H > H_c$, there is
  only one minimum, the magnetization is reversed.}
\end{figure}

In the past, magnetization reversal is realized by applying a magnetic
field pulse mainly antiparallel to the initial magnetization.  The
magnetization will then undergo multiple rotations around the local
effective field to reach the final equilibrium direction.  This
process is best viewed by drawing the energy landscape of the system.
Assume magnetization in the $\hat{x}$ and $-\hat{x}$ directions gives the
local energy minima.  As shown in Fig.~\ref{fig:landscape}, upon application
of the field pulse, the landscape is tilted, making the barrier
disappear.  The system will jump to the global minimum and the
magnetization is therefore switched from one direction to another.  It
is clear that the applied field has to be strong enough to overcome
the energy barrier.  The critical switching field is first discussed
in the uniaxial anisotropy case in the pioneering work by Stoner and
Wahlfarth,\cite{stoner1948} and recently extended to non-uniaxial and
three dimensional cases.  \cite{thiaville1998, thiaville2000} Energy
dissipation is necessary in this process for the system to move from
one minimum to another.  We call this type of reversal the
Stoner-Wahlfarth (SW) type.  Typical reversal time for such a process
is of the order of nanoseconds.

Recently, a novel approach towards ultrafast magnetization switching
by precessional motion has been proposed and observed experimentally.
\cite{back1998, back1999, bauer2000, acremann2001, miltat2001,
schumacher2003, schumacher2003a} Instead of applying a field pulse
antiparallel to the initial magnetization, a perpendicular field pulse
is applied and induces a large angle precession.  If the pulse is
terminated at 180$^\circ$ angle of the precession, whose period is
usually of the order of picoseconds, the magnetization is reversed.
However, the effective reversal times could be several nanoseconds due
to the decay time of residual magnetic precession (``ringing'') if the
magnetization does not end up in the final equilibrium state
precisely.  One can overcome this difficulty by fine tuning the pulse
parameters so that the magnetization will move along the so-called
ballistic trajectory, eliminating the ringing effect.
\cite{miltat2001} Thereby the fundamental ultrafast limit of
field induced magnetization reversal is reached.  In contrast to the
SW type reversal, the precessional switching is so fast that
dissipative effects can be neglected.  In other words, the system
energy is \emph{conserved} and the problem can be formulated
using the Hamiltonian equations.

Such a precessional switching opens a way to reduce not only the
reversal time but also the field required for switching.  As pointed
out by early numerical calculations,\cite{he1994, he1996} precessional
magnetization switching can be observed well below the Stoner-Wohlfart
limit.  This result is recently recovered by considering the
precessional switching as a result of bifurcation, the long-term
behavior of the dissipative system. \cite{acremann2001} However, to
our knowledge, no systematic study of the minimal field strength
exists in the literature and most of the work was done by numerical
means.

In this paper we investigate the minimal field
strength required for precessional switching of a
magnetic object with biaxial anisotropy.  We use phase space analysis
as our main tool to tackle this problem.  It turns out that the
mechanism of the precessional switching is directly related to the
evolution of the localized and unlocalized trajectories in the phase
space.  Unlike the SW type switching, where the fixed points and their
movement are the central objects to study, in the precessional
switching the states move along equienergy curves and we need follow
their motion globally.  The minimal (critical) field is obtained when
the localized trajectories become unlocalized.  By studying the phase
space evolution we are able to obtain the analytical expression for the
critical field.  This result provides useful information on the design
of field pulses for precessional switching based devices.

The paper is organized as follows.  In Sec.~\ref{sec:mod} we present a
simple model describing magnetization dynamics and define the central
problem.  We then in Sec.~\ref{sec:nodamping} consider the zero
damping situation in which the switching field is applied
perpendicular to the easy axis.  We also calculate the switching times
for this configuration.  In Sec.~\ref{sec:damping} with the assumption
of very small damping we study the applied field perpendicular to the
hard axis configuration and compare our result with the standard SW
model.  Finally the paper is summarized in Sec.~\ref{sec:sum}.

\section{\label{sec:mod}Model}

The simplest micromagnetic model is the so-called macro-spin model,
i.e., the magnetization is uniform and displays collective dynamics.
This model is valid for small object size (reaching the single domain
limit) and low magnetic coercivity.  The motion of magnetization
$\vect{M}$ with a phenomenological damping is governed by the
Landau-Lifshitz-Gilbert (LLG) equation.  This equation will be used in
the following dimensionless form:
\begin{equation}  \label{eq:llg}
  \ddif{\vect{m}}{t} = -\vect{m} \times \vect{h}_\eff
    + \alpha \vect{m} \times \ddif{\vect{m}}{t} \;.
\end{equation}
Here $\vect{m} = \vect{M}/M_s$ is the magnetization unit vector,
$\vect{h}_\eff = \vect{H}_\eff/M_s$ is the scaled effective field,
time is measured in units of $(\gamma M_s)^{-1}$, $M_s$ is the
saturation magnetization, $\gamma$ is the absolute value of the
gyromagnetic ratio, and $\alpha$ is the dimensionless damping
parameter.

We consider a very thin film in the $x$-$y$ plane with an in-plane
uniaxial anisotropy, taking $x$ the easy axis of magnetization.  The
demagnetizing field factors are practically equal to zero in the film
plane and one perpendicular to the film plane, respectively.  The
magnetic energy density is written
\begin{equation}
  w(\vect{m}, \vect{h}) = \frac{1}{2} m_z^2 - \frac{1}{2}K m_x^2 -
  \vect{h}\cdot\vect{m} \;,
\end{equation}
where $K>0$ accounts for the scaled in-plane $x$-axis anisotropy )and
$\vect{h}$ denotes applied field.  Usually the value of $K$ is about
0.01 for thin films because of the large demagnetizing field in the
$\hat{z}$ direction.  However, after proper scaling the above
expression can be also used to describe the energy density of small
magnetic particles with biaxial anisotropy as well.  In following
discussions we assume $K>0$.

Let us consider the energy dissipation rate.  Recall that $\vect{h}_\eff
= -\partial w/\partial \vect{m}$, we find out the dissipation rate has
the following form
\begin{equation}
  \ddif{w}{t} = \dpar{w}{\vect{m}}\cdot\ddif{\vect{m}}{t}
    = - \frac{\alpha}{1+\alpha^2}\abs{\vect{h}_\eff \times \vect{m}}^2\;.
\end{equation}
For small damping parameter $\alpha$, short pulse period, or small
applied field, the damping term in the LLG equation~\eqref{eq:llg} can
be neglected.  We note that in real experiments the
damping parameter is usually of the order of $10^{-2}$, and the field
pulse duration can be adjusted to hundreds of picoseconds, thus
providing a very good examples of this \emph{zero} damping model.  The
dynamics is described by the following equation:
\begin{equation}
  \dpar{\vect{m}}{t} = -\vect{m} \times \vect{h}_\eff \;.
\end{equation}
In terms of the polar angle $\theta$ and the azimuthal angle $\phi$,
the energy density reads
\begin{equation}
\begin{split}
  w &= \frac{1}{2}\cos^2\theta - \frac{1}{2}K\sin^2\theta\cos^2\phi \\ 
    &\quad- h_x\sin\theta\cos\phi - h_y\sin\theta\sin\phi - h_z\cos\theta \;,
\end{split}
\end{equation}
and the equations of motion are
\begin{equation}  \label{eq:eom}
  \dot{\theta} = -\frac{1}{\sin\theta}\dpar{w}{\phi} \;,\qquad 
  \dot{\phi} = \frac{1}{\sin\theta}\dpar{w}{\theta} \;.
\end{equation}

For simplicity we consider field pulses with zero rise and fall times.
That is, the applied field remains constant for a pulse
duration of $T$.  We then define the problem as follows.  Assume
initially the magnetization is along the $+\hat{x}$ direction at
$\theta=\pi/2$, $\phi=0$, denoted by $\vect{m}_0$.  By applying a
field pulse we expect the magnetization to switch to the $-\hat{x}$
direction at $\theta=\pi/2$, $\phi=\pi$, denoted by $\vect{m}_1$.  In
order to achieve this both field strength and pulse duration have to
satisfy certain requirement.  Here we will mainly consider the field
strength requirement.

\section{\label{sec:nodamping}Zero Damping: Switching Field in $y$-$z$ Plane}

We first study the $\alpha = 0$ case.  In this case,
the magnetization cannot relax to a local minimum by dissipating
energy, therefore the initial state $\vect{m}_0$ and the final state
$\vect{m}_1$ must be connected by a constant energy curve during the
application of the field pulse.  This requirement leads to $\vect{h}
\cdot (\vect{m}_1 - \vect{m}_0) = 0$; the applied field must be in the
$y$-$z$ plane.

\begin{figure}[t]
\includegraphics[width=8.5cm]{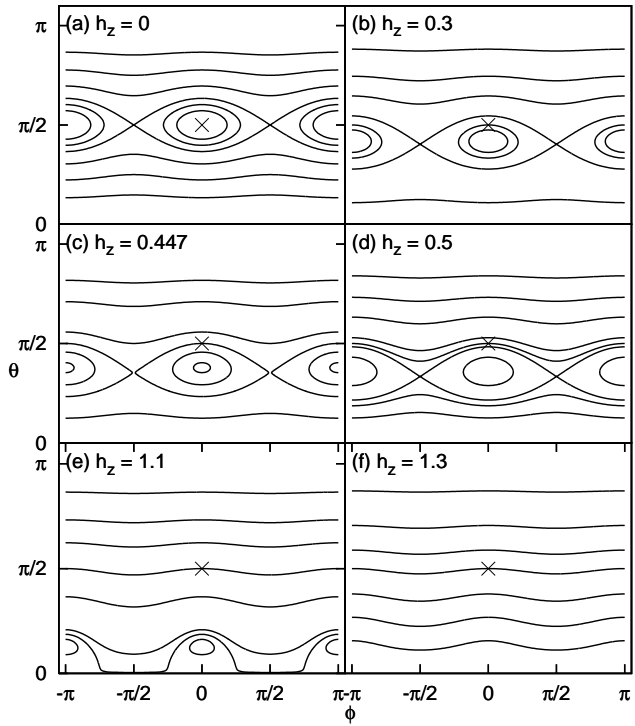}
\caption{\label{fig:phase_hz}The phase space at $K=0.2$ for different
  $h_z$.  Point X denotes the initial state $\vect{m}_0$. In (a) $h_z
  = 0$, $\vect{m}_0$ is a local minimum.  For $0 < h_z < 1$,
  $\vect{m}_0$ gradually moves out from the localized region as shown
  in (b)-(d).  In (e) $1 < h_z < 1 + K$, the two saddle points collide
  at the north pole, end up with only one saddle point.  In (f) all
  trajectories are unlocalized in the range $h_z > 1 + K$.}
\end{figure}

The phase space without applied field is shown in
Fig.~\ref{fig:phase_hz}a.  The system has (a) two minima at
$\theta=\pi/2$, $\phi = 0, \pi$, (b) two maxima at $\theta=0, \pi$
($\phi$ has no definition), and (c) two saddle points at
$\theta=\pi/2$, $\phi = \pm\pi/2$.  The phase space is divided by the
separatrix (trajectories that join the saddle points) into four
regions.  Inside the separatrix there are two regions, where the
motion is periodic and localized around the corresponding minima.  In
a dissipative system these two regions are called basin of attraction
for these two minima.  Outside the separatrix, in the top and bottom,
the motion is unlocalized, i.e., the magnetization can move from the
$+\hat{x}$ direction to the $-\hat{x}$ direction.  Generally speaking,
in a precessional switching the function of the applied field is to
tilt the phase space structure and drag the initial state $\vect{m}_0$
out of the localized region so the magnetization can precess globally.

To better illustrate this idea, we consider two special cases in which
the applied field is along the $\hat{z}$ direction and $\hat{y}$
direction, respectively.

In case (i), the field $\vect{h}$ is applied along the $\hat{z}$
direction.  The phase space at $K < 1$ is shown in
Fig.~\ref{fig:phase_hz}b-f.  In the range of $h_z < 1$, the phase
space has the same topological structure as $h_z=0$.  There are (a)
two minima at $\theta=\arccos[h_z/(1+K)]$, $\phi=0, \pi$, (b) two
maxima at $\theta = 0, \pi$, and (c) two saddle points at
$\theta=\arccos h_z$, $\phi=\pm\pi/2$.  In the range $0 < h_z <
\sqrt{K}$, the initial state $\vect{m}_0$ stays in the central
localized region, no magnetization switching will occur
(Fig.~\ref{fig:phase_hz}b).  If $h_z = \sqrt{K}$, $\vect{m}_0$ is on
the separatrix and it could evolve into the other localized region,
that is when the switching starts (Fig.~\ref{fig:phase_hz}c).  Note
that on the separatrix the motion is however very slow when
approaching the saddle points..  As shown in Fig.~\ref{fig:phase_hz}e, in the range $1 <
h_z < 1 + K$, the two saddle points at $\theta=\arccos h_z$,
$\phi=\pm\pi/2$ disappear, and $\theta=0$ becomes a saddle
point, i.e., the separatrix passes the north pole.  Finally for $h_z > 1+K$ there are only one minimum at
$\theta=0$ and one maximum at $\theta=\pi$, all trajectories are
unlocalized.

\begin{figure}[t]
\includegraphics[width=8.5cm]{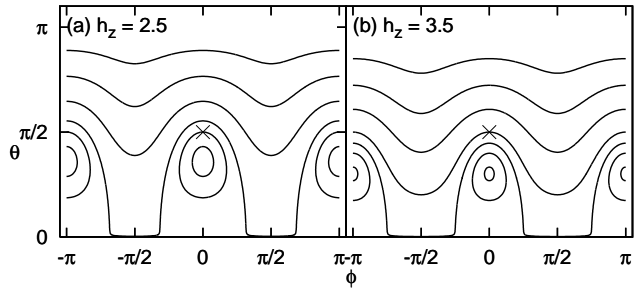}
\caption{\label{fig:phase_hz1}The phase space at $K=5$ for different
  $h_z$.  Point X denotes initial state $\vect{m}_0$.  In (a) $1 < h_z
  < (1+K)/2$, $\vect{m}_0$ stays in the localized region.  In (b)
  $(1+K)/2 < h_z < 1+K$, $\vect{m}_0$ is on a unlocalized trajectory.}
\end{figure}

If $K>1$, i.e., the anisotropy of the easy axis is bigger than that of
the hard axis, the two wells around the energy minima are so deep that
even after $h_z$ passes 1 the initial state $\vect{m}_0$ still stays
in the localized region, shown in Fig.~\ref{fig:phase_hz1}a.
calculation shows the minimal field strength that causes switching is
$h_z = (1+K)/2$, bigger than $\sqrt{K}$.

\begin{figure}[t]
\includegraphics[width=8.5cm]{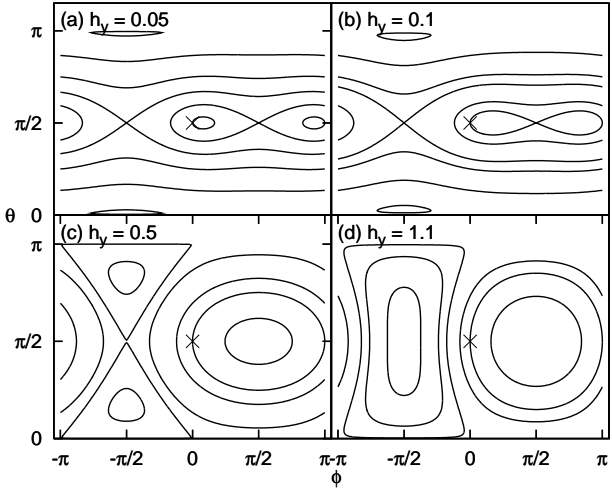}
\caption{\label{fig:phase_hy} The phase space at $K=0.2$ for different
  $h_y$.  Point X denotes the initial state $\vect{m}_0$. In (a) and
  (b) $0 < h_y < K$.  At $h_y = K/2$, $\vect{m}_0$ is on the
  separatrix.  In (c) $K < h_y < 1$.  In (d) $h_y > 1$.}
\end{figure}

Now let us consider the case where the field $\vect{h}$ is applied
along the $hat{y}$ direction.  The $h_y$ field breaks the $y$ axis
symmetry.  As a consequence, the two saddle points have different
energies.  The phase space at $K<1$ is shown in
Fig.~\ref{fig:phase_hy}.  In the range of $0 < h_y < K$, there are (a)
two minima at $\theta=\pi/2$, $\phi=\arcsin h_y/K, \pi-\arcsin h_y/K$,
(b) two maxima at $\theta = \arcsin h_y, \pi - \arcsin h_y$, $\phi =
-\pi/2$, and (c) two saddle points at $\theta=\pi/2$, $\phi=\pm\pi/2$.
As $h_y$ increases, the size of localized regions shrinks.  The
precessional switching occurs at $h_y = K/2$
(Fig.~\ref{fig:phase_hy}b).  For$h_y > K$, the two minima
collide together with one of the saddle points and disappear, a new
minima emerges at $\theta=\pi/2$, $\phi=\pi/2$.  After $h_y$ passes 1
there are only one minimum at $\theta=\pi/2$, $\phi=\pi/2$ and one
maximum at $\theta=\pi/2$, $\phi=-\pi/2$.  If $K>1$, evolution of the
phase space will be different, but the critical condition $h_y = K/2$
still holds.

From the above discussion, we can see that precessional switching
occurs during the transition of trajectories around the initial state
$\vect{m}_0$ from localized to unlocalized.  In other words, this is
when $\vect{m}_0$ moves from inside to outside the separatrix, so the
critical field is the one that makes $\vect{m}_0$ on the separatrix.
Since points in the phase space move along trajectories of constant
energy, the critical condition can be also stated as that the energy
of initial state equals the energy of saddle points (but we need check
if they are on the same trajectory).  We observe that this situation
usually happens before collision of two fixed points, which is the
critical condition for the SW type reversal.  This explains the
smaller switching field in precessional switching.

Now we consider the general case in which the field $\vect{h}$ is
applied in the $y$-$z$ plane at an arbitrary angle.  The energies of
the initial state $\vect{m}_0$ and final state $\vect{m}_1$ are always
$-K/2$.  We are looking for saddle points with the same energy.  The
fixed points are given by
\begin{subequations}
\begin{gather}
  -\sin\theta\cos\theta(1+K\cos^2\phi) - h_y\cos\theta\sin\phi 
       + h_z\sin\theta = 0 \;,  
  \label{eq:fp1} \\
  K\sin^2\theta\sin\phi\cos\phi - h_y\sin\theta\cos\phi = 0 \;.
  \label{eq:fp2}
\end{gather}
\end{subequations}
From Eq.~\eqref{eq:fp2} we see that solutions can be grouped into two
categories.  The first category includes solutions with $\sin\phi =
h_y/(K\sin\theta)$.  Inserting this expression into Eq.~\eqref{eq:fp1}
and solving it gives us fixed points
\begin{equation}  \label{eq:minima}
  \cos\theta = \frac{h_z}{1+K} \;, \quad
  \sin\phi = \frac{h_y}{K\sqrt{1-h_z^2/(1+K)^2}} \;,
\end{equation}
with the condition
\begin{equation}  \label{eq:cond}
  \frac{h_y^2}{K^2} + \frac{h_z^2}{(1+K)^2} \le 1 \;.
\end{equation}
In following discussion we assume this condition always holds since we
are interested in the minimal field strength.  We will see later the
result is consistent with this assumption.  The energy of the fixed
points is
\begin{equation}
  E = -\frac{K}{2} 
    - \frac{1}{2}\Bigl(\frac{h_y^2}{K} + \frac{h_z^2}{1+K}\Bigr) \;.
\end{equation}
The energy is always smaller than $-K/2$ so they are not what we are
looking for.  Stability analysis shows in fact they are energy minima.

Now we turn to the other category, which includes solutions with
$\cos\phi = 0$.  In this category, fixed points occur on a great
circle parametrized by setting $\phi=0$ and letting $\theta$ run from
0 to $2\pi$.  This makes the problem essentially a two-dimensional
problem in the $y$-$z$ plane.
\begin{figure}[t]
\includegraphics[width=8cm]{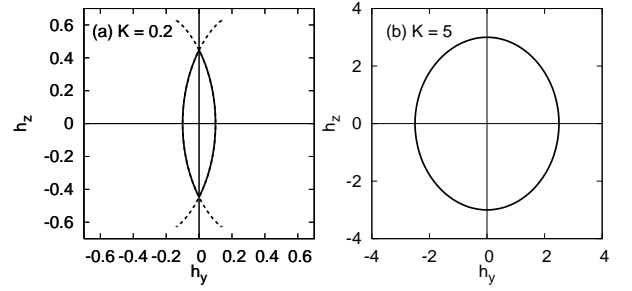}
\caption{\label{fig:field_yz}Critical switching field in the $y$-$z$
  plane at $\alpha = 0$ for (a) $K<1$ and (b) $K>1$.  The
  intersections of the curve are in (a) $h_y = K/2$, $h_z = \sqrt{K}$,
  in (b) $h_y = K/2$, $h_z = (K+1)/2$.  Fields shown in dashed line
  corresponds to the situation where the initial state has the same
  energy with one of the saddle points but is not on the separatrix.}
\end{figure}

Instead of numerics, this problem can be solved by a geometrical
approach, first proposed by Slonczewski in the Stoner-Wohlfarth
problem, then extended by Thiaville. \cite{thiaville1998,
thiaville2000} In essence, it considers the field $\vect{h}$, instead
of the magnetization direction $\vect{m}$ as the main variable.  We
write $\vect{m} = \sin\theta\hat{y} + \cos\theta\hat{z} = (\sin\theta,
\cos\theta)$ and its orthogonal vector $\vect{e} = (\cos\theta,
-\sin\theta)$.  The energy requirement is
\begin{equation}  \label{eq:energy}
  E = \frac{1}{2}\cos^2\theta - \vect{h}\cdot\vect{m} = -\frac{K}{2} \;.
\end{equation}
The extremum condition is
\begin{equation}  \label{eq:extremum}
  \ddif{E}{\theta} = -\sin\theta\cos\theta - \vect{h}\cdot\vect{e} = 0 \;.
\end{equation}
Since the system already has two minima at points~\eqref{eq:minima}, the saddle
point we are looking for has to be the minimum along the
$\hat{\theta}$ direction.  This leads to
\begin{equation}  \label{eq:saddle}
  \ddif{^2E}{\theta^2} = \sin^2\theta - \cos^2\theta +
  \vect{h}\cdot\vect{m} > 0 \;.
\end{equation}

The critical switching field is determined by the intersections of the
two lines defined in Eqs.~\eqref{eq:energy} and \eqref{eq:extremum}.
We find
\begin{subequations}
\begin{align}
    h_y &= \frac{1}{2}\sin\theta(K - \cos^2\theta) \;, \\
    h_z &= \frac{1}{2}\cos\theta(1 + K + \sin^2\theta) \;.
\end{align}
\end{subequations}
Inserting above expressions into Eq.~\eqref{eq:saddle} yields
\begin{equation}
  \ddif{^2E}{\theta^2} = 1 + \frac{1}{2}K - \frac{3}{2}\cos^2\theta
     > 0 \;.  
\end{equation}
For $K > 1$, the above inequality always holds.  For $K < 1$ we have
$\abs{\cos\theta} < \sqrt{(2+K)/3}$.  However, in this range the
critical curve does not form a closed loop, shown in
Fig.~\ref{fig:field_yz}a.  The reason is that at nonzero $h_y$ the two
saddle points have different energies so the trajectory starting from
$\vect{m}_0$ may not pass that saddle point.  After checking we can
remove these ``faked'' solutions by requiring $\abs{\cos\theta} <
\sqrt{K}$.

\begin{figure}[t]
\includegraphics[width=8cm]{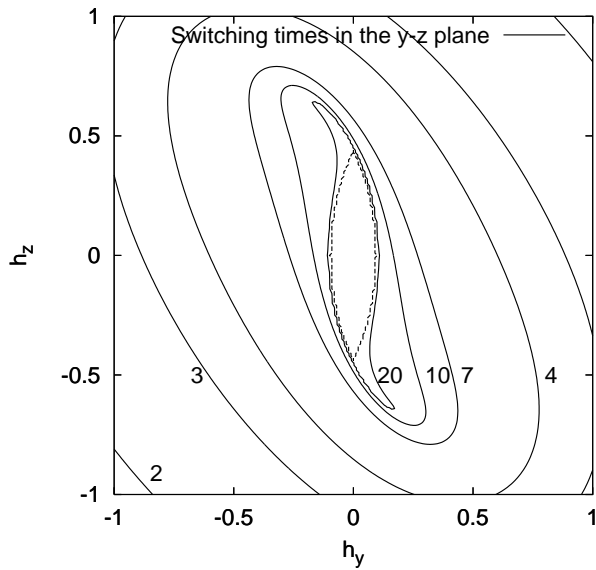}
\caption{\label{fig:times}Switching times contour in the $y$-$z$ plane
  for $K=0.2$.  The dashed line in the center is the critical
  switching field.  From outside to inside, the switching times are 2,
  3, 4, 7, 10, 20, measured in $(\gamma M_s)^{-1}$.  For typical
  configuration it is about 0.01 ns.}
\end{figure}

From Fig.~\ref{fig:field_yz}
we can see that if $K \gg 1$ the magnitude of critical switching field
along any direction in the $y$-$z$ plane is about the same; for
magnetic thin films where $K \ll 1$ then $\hat{y}$ direction is the
best choice to apply small field.  We also calculate the switching
time for $K<1$ by numerically integrating Eq.~\eqref{eq:eom}.  The
equi-time contours are plotted in Fig.~\ref{fig:times}.  As the
magnitude of fields increases, the switching time decreases quickly
from the center.

\section{\label{sec:damping}Small Damping: Switching Field in $x$-$y$ Plane}

In this section we study the small damping case.  As finding the exact
analytical solution to the equation of magnetization motion including
damping is a formidable task, we turn to study the limiting case,
where $\alpha$ is small enough so that we can still treat the problem
as a Hamiltonian problem and $\alpha$ is also big enough so that after
the application of the field pulse the magnetization can relax to the
local minimum in finite time.  Because of damping the energies of the
initial and final state need not to be the same, a nonzero field
component along the $\hat{x}$ direction is allowed.

We choose to apply the field in the $x$-$y$ plane.  By this way, we
can compare our result with the SW model.  The basic procedure is the
same as last section: we will search the saddle points with the same
energy of the initial state, then map to the whole switching field.

It can be seen from Fig.~\ref{fig:phase_hy} that the two fixed points
that originate from the south and north pole do not involve in this
precessional switching, so the saddle point we are looking for
is on the $x$-$y$ plane too.  Let $\theta=\pi/2$, then the problem
turns into a two-dimensional problem in the $x$-$y$ plane.  We write
$\vect{m} = \cos\phi\hat{x} + \sin\phi\hat{y} = (\cos\phi, \sin\phi)$
and its orthogonal vector $\vect{n} = (-\sin\phi, \cos\phi)$.  The
applied field is $\vect{h} = (h_x, h_y)$.

\begin{figure}[t]
\includegraphics[width=8cm]{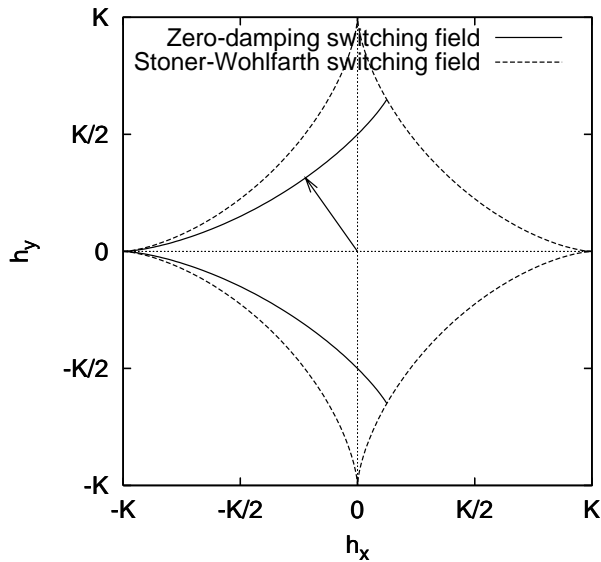}
\caption{\label{fig:field_xy}Critical switching field in the $x$-$y$
  plane for switching from the $\hat{x}$ direction to the $-\hat{x}$
  direction.  The solid line shows the switching field of precessional
  switching at zero damping limit.  The dashed line is the switching field
  of the Stoner-Wohlfarth model.  For completion we also draw the
  right half of the SW astroid.  Arrow indicates the minimal
  field strength.}
\end{figure}

The energy requirement is
\begin{equation}
  E = -\frac{K}{2}\cos^2\phi - \vect{h}\cdot\vect{m}
  = -\frac{K}{2} - h_x \;.
\end{equation}
The extremum condition is
\begin{equation}
  \ddif{E}{\phi} = K\cos\phi\sin\phi - \vect{h}\cdot\vect{n} = 0 \;.
\end{equation}
Solving above equations gives us the switching field in the $x$-$y$
plane:
\begin{subequations}
\begin{align}
  h_x &= -\frac{K}{2}\cos\phi(1+\cos\phi) \;, \\ 
  h_y &= \frac{K}{2}\sin\phi(1-\cos\phi) \;.
\end{align}
\end{subequations}
In order to check the stability of the corresponding fixed point we
carry out the standard routine.  With $h_x$, $h_y$ taking the above
form, the second derivatives of $w$ are
\begin{subequations}
\begin{align}
  \dpar{^2w}{\theta^2} &= 1 + \frac{K}{2}(1-\cos\phi) \;, \\
  \dpar{^2w}{\phi^2} &= K(\cos\phi-1)(\cos\phi+\frac{1}{2}) \;, \\
  \dpar{^2w}{\theta\partial\phi} &= 0 \;.
\end{align}
\end{subequations}
It is obvious that $\phi$ has to in the range $0< \phi <
\frac{3}{4}\pi$ (upper branch in Fig.~\ref{fig:field_xy}) or
$-\frac{3}{4}\pi < \phi < 0$ (lower branch in Fig.~\ref{fig:field_xy})
to make the fixed point a saddle point.  Again we still need to check
if the saddle point and the initial state are on the same trajectory.
We require both $h_x$ and $h_y$ are monotonous function of $\phi$ in
the upper and lower branches, which narrows the range to
$-\frac{2}{3}\pi < \phi < \frac{2}{3}\pi$.

To compare with the switching field in the SW model, which takes the
form $\vect{h} = (-\cos^3\phi, \sin^3\phi)$, we plot both fields in
Fig.~\ref{fig:field_xy}.  It is clear that precessional switching can
occur well below the SW limit.  Surprisingly, magnetic switching is
possible for a field with a positive $x$ component.  Although in this
case the minimal field strength is $0.38K$ ($cos\phi=1/3$),
considering the long relaxation time it is still better to apply
the field mainly along $\hat{y}$ direction with a small $x$ component
to compensate the energy loss during the motion.

\section{\label{sec:sum}Summary}

The critical switching field of a monodomain magnetic object with
biaxial anisotropy has been studied.  In particular, switching fields
perpendicular to the easy and hard axes are calculated using the
geometrical method.  It is found that in the limit of zero damping,
applying field along the medium axis is good for both small fields
and fast switching times.

This method can be generalized for magnetic particles with a higher
order anisotropy if one follows the same steps outlined in this paper.
The two-dimensional model may not be sufficient because one needs to
consider more fixed points.  A more complicated three-dimensional
model would then be required.

\begin{acknowledgments}
We thank the NSF NIRT program (Grant No.~DMR-0404252) for support of
this work.
\end{acknowledgments}

\bibliography{reversal}

\end{document}